# Dimension-Reduced ADP for Real-Time Microgrid Operation with Massive Air-Conditioning Loads under Multiple Uncertainties

Jingguan Liu
*State Key Laboratory of Advanced Electromagnetic Technology Huazhong University of Science and Technology*
Wuhan, China
jingguanliu@hust.edu.cn

Xiaomeng Ai
*State Key Laboratory of Advanced Electromagnetic Technology Huazhong University of Science and Technology*
Wuhan, China
xiaomengai@hust.edu.cn

Shichang Cui
*State Key Laboratory of Advanced Electromagnetic Technology Huazhong University of Science and Technology*
Wuhan, China
shichang_cui@hust.edu.cn

Jiakun Fang
*State Key Laboratory of Advanced Electromagnetic Technology Huazhong University of Science and Technology*
Wuhan, China
jfa@hust.edu.cn

Jinyu Wen
*State Key Laboratory of Advanced Electromagnetic Technology Huazhong University of Science and Technology*
Wuhan, China
jinyu.wen@hust.edu.cn

***Abstract*—This paper proposes a dimension-reduced approximate dynamic programming (ADP) method for real-time microgrid operation with massive air-conditioning loads under multiple uncertainties. The operation problem is formulated as a multi-stage Markov decision process, and a post-decision value function is introduced to characterize the impact of current decisions on future operating costs. To address the curse of dimensionality caused by massive air-conditioning loads, a consistency-based value function projection is developed to map the high-dimensional state space at each node into a tractable aggregated state space. Based on the reduced states, piecewise linear approximation is further employed for efficient value function training. Case studies on 33-bus and 123-bus systems show that the proposed method achieves near-optimal operation performance with low computational cost and good scalability under both deterministic and stochastic conditions.**

## NOMENCLATURE

### A. Indices and Sets

| | |
|---|---|
| $\tau$ | Time instant. |
| $t$ | Period index. |
| $i/j$ | Bus index. |
| $n$ | Air-conditioning index. |
| $N(i)$ | Air-conditioning set at bus $i$. |
| $s$ | PLF segment index. |

### B. Parameters

| | |
|---|---|
| $\overline{[\cdot]}/\underline{[\cdot]}$ | Maximum/Minimum value of $[\cdot]$. |
| $\Delta t$ | Length of each period. |
| $T^{out}$ | Outdoor temperature. |
| $C/R$ | Thermal capacity/resistance of rooms. |
| $Q^d$ | Heat dissipation of rooms. |
| $T^{set}/\varepsilon$ | Indoor temperature set-point/tolerance. |
| $P^{ac,B}$ | Baseline power of air conditioners. |
| $\alpha/\beta/\gamma/k/l$ | Thermoelectric coefficients. |
| $c^{buy}/c^{sell}$ | Cost coefficient of energy purchasing/selling. |
| $c^{gt}$ | Cost coefficient of gas turbines. |
| $c^{dr}$ | Cost coefficient of air conditioner response. |
| $c^{wc}$ | Cost coefficient of wind curtailment. |
| $\varphi$ | Power factor angle of air conditioners. |
| $G/B$ | Branch conductance/susceptance matrices. |
| $g/b$ | Branch conductance/susceptance. |
| $B'$ | Branch susceptance matrices without shunt elements. |
| $P^{wd}/P^{ld}$ | Forecast wind and load active power. |
| $Q^{ld}$ | Forecast load reactive power. |

### C. Decision Variables

| | |
|---|---|
| $T^{in}/Q$ | Indoor temperature/cooling capacity. |
| $P^{ac}/Q^{ac}$ | Active/Reactive power of air conditioners. |
| $P^{buy}/P^{sell}$ | Active power of purchasing/selling. |
| $P^{gt}/Q^{gt}$ | Active/Reactive power of gas turbines. |
| $P^{mg}/Q^{mg}$ | External active/reactive power injection. |
| $P^{dr}$ | Active power of air conditioner response. |
| $P^{wc}$ | Active power of wind curtailment. |
| $V/\theta$ | Bus voltage magnitude/phase angle. |
| $P^L$ | Active power flow. |

## I. INTRODUCTION

The increasing penetration of renewable energy sources in microgrids poses significant challenges to real-time operation because of their intermittency and uncertainty [1], which require greater operational flexibility to maintain the supply-demand balance [2]. Air-conditioning loads, whose thermal inertia enables short-term power adjustment without significantly affecting user comfort, have emerged as a promising source of flexibility [3]. With the growing integration of air-conditioning loads and renewable generation, microgrids can actively coordinate these flexible resources to achieve more economic real-time operation [4].

A key challenge in real-time microgrid operation lies in its sequential nature [5]. Uncertainties in wind power, load demand, and outdoor temperature are gradually revealed over time, and decisions at each period must be made based only on currently observed information [6]. Conventional approaches, such as robust optimization [7], stochastic programming [8], and model predictive control [9], either do not explicitly capture this sequential decision process or become computationally burdensome when handling multiple uncertainties. Although multi-stage formulations can better reflect the temporal structure of the problem, their computational complexity grows rapidly with problem dimension [10].

Approximate dynamic programming (ADP) provides a principled framework for sequential decision-making under uncertainty by decomposing the original multi-stage problem into a series of time-dependent subproblems based on Bellman's equation [11]. The value function, trained offline using historical data, captures the impact of current decisions on future operating costs and thus supports near-optimal real-time operation [10]. However, existing ADP methods are typically developed for systems with only a limited number of storage-like resources, where the value function space remains manageable. In microgrids with massive air-conditioning loads, the value function space becomes extremely high-dimensional, making both value function approximation and offline training computationally intractable.

To address this challenge, this paper proposes a dimension-reduced ADP method for real-time microgrid operation with massive air-conditioning loads under multiple uncertainties. The main idea is a consistency-based value function projection that maps the high-dimensional value function space of air-conditioning loads at each node into a tractable aggregated state space. This reformulation preserves the sequential structure of the original problem while substantially reducing the computational burden, thereby enabling efficient piecewise linear approximation and temporal-difference learning. Case studies on the 33-bus and 123-bus systems demonstrate that the proposed method achieves near-optimal operation performance with low computational cost and good scalability.

The remainder of this paper is organized as follows. Section II presents the multi-stage real-time operation formulation. Section III describes the proposed dimension-reduced ADP method. Section IV presents the case studies. Section V concludes the paper.

## II. Problem Formulation

This section presents the thermal dynamics of air-conditioning loads and incorporates them into the real-time operation model of a microgrid.

### A. Thermal Dynamics of Air-Conditioning Loads

The thermal behavior of an air-conditioned room is modeled by a first-order equivalent thermal parameter model [12]. The continuous-time evolution of indoor temperature is given by

$$CdT^{in}(\tau)/d\tau = (T^{out}(\tau) - T^{in}(\tau))/R + Q^d - Q(\tau) \quad (1)$$

Assuming that the cooling power and outdoor temperature remain constant over a short interval $\Delta t$ within $[t-1, t]$, (1) can be discretized as

$$T_t^{in} = T_t^{out} + R(Q^d - Q_t) - [T_t^{out} + R(Q^d - Q_t) - T_{t-1}^{in}]e^{-\Delta t/(RC)} \quad (2)$$

In addition, the electro-thermal coupling of air-conditioning loads is described by (3), and the corresponding baseline power is given by (4).

$$P_t^{ac} = kQ_t + l \quad (3)$$

$$P_t^{ac,B} = k(T_t^{out} - T^{set})/R + l \quad (4)$$

### B. Real-Time Operation Formulation

In this paper, the real-time operation problem of a microgrid considers distributed gas turbines, the external grid, wind turbines, buildings with air-conditioning loads, and rigid loads. The corresponding formulation is given in (5)-(12).

$$\min \mathrm{E}\left\{\sum_t C_t^{obj}\right\},$$
$$C_t^{obj} = \sum_i \begin{pmatrix} (c_{i,t}^{buy} P_{i,t}^{buy} - c_{i,t}^{sell} P_{i,t}^{sell}) + c_{i,t}^{gt} P_{i,t}^{gt} \\ + \sum_{n\in N(i)} (c_{i,n,t}^{dr} P_{i,n,t}^{dr}) + c_{i,t}^{wc} P_{i,t}^{wc} \end{pmatrix} \quad (5)$$

$$\textbf{s.t.} \begin{cases} \text{Constraints (2)-(4)} \\ T_{i,n}^{set} - \varepsilon_{i,n} \le T_{i,n,t}^{in} \le T_{i,n}^{set} + \varepsilon_{i,n} \end{cases} \quad (6)$$

$$\begin{cases} Q_{i,n,t}^{ac} = P_{i,n,t}^{ac} \tan\varphi_{i,n}, \underline{P_{i,n}^{ac}} \le P_{i,n,t}^{ac} \le \overline{P_{i,n}^{ac}} \\ P_{i,n,t}^{dr} \ge P_{i,n,t}^{ac} - P_{i,n,t}^{ac,B}, P_{i,n,t}^{dr} \ge P_{i,n,t}^{ac,B} - P_{i,n,t}^{ac} \end{cases} \quad (7)$$

$$P_{i,t}^{mg} + P_{i,t}^{gt} + P_{i,t}^{wd} - P_{i,t}^{wc} - P_{i,t}^{ld} - \sum_{n\in N(i)} P_{i,n,t}^{ac} = \sum_j G_{ij} V_{j,t} + \sum_j B'_{ij} \theta_{j,t} \quad (8)$$

$$Q_{i,t}^{mg} + Q_{i,t}^{gt} - Q_{i,t}^{ld} - \sum_{n\in N(i)} Q_{i,n,t}^{ac} = -\sum_j B_{ij} V_{j,t} - \sum_j G_{ij} \theta_{j,t} \quad (9)$$

$$\begin{cases} \underline{V_i} \le V_{i,t} \le \overline{V_i}, -\pi \le \theta_{i,t} \le \pi \\ P_{ij,t}^L = g_{ij}(V_{i,t} - V_{j,t}) - b_{ij}(\theta_{i,t} - \theta_{j,t}), -\overline{P_{ij}^L} \le P_{ij,t}^L \le \overline{P_{ij}^L} \end{cases} \quad (10)$$

$$\underline{P_i^{gt}} \le P_{i,t}^{gt} \le \overline{P_i^{gt}}, \underline{Q_i^{gt}} \le Q_{i,t}^{gt} \le \overline{Q_i^{gt}}, 0 \le P_{i,t}^{wc} \le P_{i,t}^{wd} \quad (11)$$

$$\begin{cases} P_{i,t}^{mg} = P_{i,t}^{buy} - P_{i,t}^{sell}, P_{i,t}^{buy} \ge 0, P_{i,t}^{sell} \ge 0 \\ -\overline{P_i^{mg}} \le P_{i,t}^{mg} \le \overline{P_i^{mg}}, -\overline{Q_i^{mg}} \le Q_{i,t}^{mg} \le \overline{Q_i^{mg}} \end{cases} \quad (12)$$

In (5), the objective is to minimize the expected operating cost over the operation horizon under uncertainties in wind power, load demand, and outdoor temperature. As shown in the second line of (5), the total cost consists of the power exchange cost, the generation cost of gas turbines, the response cost of air-conditioning loads, and the penalty cost of wind curtailment. Constraints (6) and (7) define the power and temperature limits of air-conditioning loads. Constraints (8) and (9) represent decoupled linearized power flow equations that account for both active and reactive power and provide high approximation accuracy with respect to the original alternating-current power flow model. Constraint (10) imposes limits on voltage magnitudes, phase angles, and branch capacities. Constraint (11) defines the power limits of gas turbines and wind curtailment. Constraint (12) specifies the power exchange limit of the external grid.

## III. Dimension-Reduced ADP Method

This section presents a dimension-reduced ADP method for efficiently solving the real-time operation problem.

### A. Reformulation Based on Bellman Optimality Equation

In real-time operation, decisions are made in a rolling manner; that is, the control action at each time step is determined according to the currently observed realizations of uncertain factors and the current system states. To explicitly capture the sequential nature of real-time microgrid operation under multiple uncertainties, the above model is reformulated as a multi-stage Markov decision process consisting of four components: state variables, decision variables, exogenous information, and transition functions [10].

The state variables describe the current system condition and contain sufficient information for decision-making. In this paper, the indoor temperatures of air-conditioning loads at the beginning of each time interval are taken as the state variables, as defined in (13).

$$S_t = \left\{ T_{i,n,t}^{in} \right\} \tag{13}$$

The decision variables represent the control actions taken after the current state is observed, as given in (14) .

$$x_t = \left\{ \begin{array}{l} P_{i,t}^{buy}, P_{i,t}^{sell}, P_{i,t}^{gt}, P_{i,n,t}^{dr}, P_{i,t}^{wc}, P_{i,n,t}^{ac}, P_{i,t}^{mg}, \\ P_{ij,t}^{L}, Q_{i,t}^{mg}, Q_{i,t}^{gt}, Q_{i,n,t}^{ac}, V_{j,t}, \theta_{j,t}, Q_{i,n,t} \end{array} \right\} \tag{14}$$

The exogenous information represents the realizations of uncertain factors, including wind power, outdoor temperature, and rigid load demand, as shown in (15).

$$\hat{W}_t = \left\{ \hat{P}_{i,t}^{wd}, \hat{T}_{i,n,t}^{out}, \hat{P}_{i,t}^{ld}, \hat{Q}_{i,t}^{ld} \right\} \tag{15}$$

The transition function describes the evolution of the system state under the decision variables and exogenous information, as shown in (16). Here, $C^{[\cdot]}$ denotes compact coefficient matrices extracted from (6)-(12), where each equality constraint can be equivalently transformed into two opposite inequalities.

$$C^{c1} S_t + C^{c2} x_t + C^{c3} S_{t-1} \le C^{c4} + C^{c5} \hat{W}_t \tag{16}$$

Based on the above elements, the globally optimal solution to the multi-period problem in (5)-(12) can be obtained by recursively solving the single-period Bellman optimality equation in (17), where the value function $V_t(S_t)$ represents the sum of the current operating cost and the expected future cost from period $t+1$ to the end of the horizon:

$$V_t(S_t) = \min \left\{ \underbrace{C_t^{obj}(S_t, x_t)}_{\text{Current cost}} + \underbrace{\mathrm{E}[V_t(S_t) \mid S_{t-1}]}_{\text{Future cost}} \right\} \tag{17}$$

In principle, (17) can be solved backward in time using classical dynamic programming. However, under stochastic environments, explicitly evaluating the expected future cost becomes computationally intractable because it requires enumerating a large number of possible realizations of the uncertainties [11]. To address this issue, the post-decision value function, denoted by $V_t^x(S_t^x)$ , is introduced to approximate the expected future cost around the post-decision state $S_t^x$ [10]:

$$V_t(S_t) = \min \left\{ C_t^{obj}(S_t, x_t) + V_t^x(S_t^x) \right\} \tag{18}$$

The post-decision state refers to the system state immediately after the decision $x_t$ is made but before the uncertainty $\widehat{W}_{t+1}$ is realized. By introducing the post-decision value function, the optimization no longer needs to explicitly enumerate a large number of uncertainty realizations, thereby significantly reducing the computational burden.

### B. Dimension-Reduced Reformulation

To obtain high-quality decisions, the post-decision value function should be accurately approximated to establish the mapping from $S_t^x$ to $V_t^x$. However, because of the massive number of air-conditioning loads, the corresponding value function space is extremely high-dimensional, making direct approximation computationally intractable. To address this issue, a consistency-based reformulation is introduced to reduce the problem dimension.

First, the state of comfort (SOC), which represents the normalized indoor temperature, is defined in (19). Its bounds are given in (20), which are derived from (6).

$$SOC_{i,n}(\tau) = (T_{i,n}^{in}(\tau) - T_{i,n}^{set}) / \varepsilon_{i,n} \tag{19}$$

$$-1 \le SOC_{i,n}(\tau) \le 1 \tag{20}$$

Based on the SOC definition, (1) can be recast as:

$$Q_{i,n}(\tau) = \alpha_{i,n} SOC_{i,n}(\tau) + \beta_{i,n} \frac{dSOC_{i,n}(\tau)}{d\tau} + \gamma_{i,n}(\tau) \tag{21}$$

where

$$\alpha_{i,n} = -\frac{\varepsilon_{i,n}}{R_{i,n}}, \beta_{i,n} = -\varepsilon_{i,n} C_{i,n}, \gamma_{i,n}(\tau) = \frac{T_{i,n}^{out}(\tau) - T_{i,n}^{set}}{R_{i,n}} + Q_{i,n}^{d}$$

To maintain SOC consistency among air-conditioning loads connected to the same node, the SOC and its derivative $dSOC/d\tau$, are enforced to be identical and are denoted by $\widehat{SOC}$ and $d\widehat{SOC}/d\tau$, respectively, as shown in (22):

$$\begin{cases} \widehat{SOC}_i(\tau) = SOC_{i,1}(\tau) = \ldots = SOC_{i,n}(\tau) \\ \dfrac{d\widehat{SOC}_i(\tau)}{d\tau} = \dfrac{dSOC_{i,1}(\tau)}{d\tau} = \ldots = \dfrac{dSOC_{i,n}(\tau)}{d\tau} \end{cases} \tag{22}$$

Accordingly, (21) can be efficiently summed over all air-conditioning loads at the same node and further recast as

$$\sum_{n \in N(i)} Q_{i,n,t} = \hat{\alpha}_i \cdot \widehat{SOC}_{i,t} + \hat{\beta}_i \cdot \left( \widehat{SOC}_{i,t+1} - \widehat{SOC}_{i,t} \right) + \hat{\gamma}_{i,t} \tag{23}$$

where

$$\hat{\alpha}_i = -\sum_{n \in N(i)} (\varepsilon_{i,n} / R_{i,n}), \hat{\gamma}_{i,t} = \sum_{n \in N(i)} ((T_{i,n,t}^{out} - T_{i,n}^{set}) / R_{i,n} + Q_{i,n}^{d})$$

$$\hat{\beta}_i = -\sum_{n \in N(i)} (\varepsilon_{i,n} / (R_{i,n}(1 - e^{-\Delta t/(R_{i,n} C_{i,n})})))$$

The state variable can then be replaced by

$$S_t = \left\{ \widehat{SOC}_{i,t} \right\} \tag{24}$$

After this reformulation, the extremely high-dimensional value function space associated with individual air-conditioning loads at each node is projected onto a single aggregated-state value function space. In this way, the main thermal dynamics are preserved while the computational burden is significantly reduced. This projection is applied only to air-conditioning loads at each node, since they play similar functional roles in microgrid operation, thereby achieving a favorable trade-off between solution quality and computational tractability.

Meanwhile, the consistent SOC signal enables automatic and fair allocation of power responses among individual air-conditioning loads. This also facilitates more practical utilization of their flexibility, since the unified signal simplifies information exchange between the system operator and the loads while reducing the risk of premature dropout of individual units when their SOCs reach operational limits.

### C. Value Function Approximation Around Reduced States

After the dimension reduction, a set of convex piecewise linear functions (PLFs), denoted by $\bar{V}_t^x(S_t^x)$, is used to approximate the post-decision value function $V_t^x(S_t^x)$ around $\widehat{SOC}$. PLFs are adopted because they provide both high approximation quality and favorable computational efficiency for storage-like devices [11]. The mathematical formulation of the PLFs is given in (25), and the problem (18) can thus be reformulated as (26)-(27).

$$V_t^x\left(S_t^x\right) \approx \bar{V}_t^x\left(S_t^x\right) = \sum_i \sum_s r_{i,t,s} \cdot v_{i,t,s} \tag{25}$$

$$V_t(S_t) = \min\left\{C_t^{obj}(S_t, x_t) + \sum_i \sum_s r_{i,t,s} \cdot v_{i,t,s}\right\} \tag{26}$$

$$\text{s.t.} \begin{cases} 0 \le v_{i,t,s} \le 2/N_s, \sum_s v_{i,t,s} - 1 = \widehat{SOC}_{i,t} \\ (3)-(5),(7)-(12),(19)-(20),(23) \end{cases} \tag{27}$$

where parameter $r_{i,t,s}$ denotes the slope of each PLF segment, and the corresponding variable $v_{i,t,s}$ represents the SOC quantity allocated to each segment. The slopes of the PLF segments can be efficiently learned from historical data using a temporal-difference learning method [13] (see Algorithm 1).

**Algorithm 1** Offline temporal-difference learning

**Step 1**: Set the number of iterations $N_M$, the number of PLF segments $N_s$, the initial slopes $r_{i,t,s,0} = 0$.
**For** $m = 1, \dots, N_M$:
/* **Forward Pass** */
**For** $t = 1, \dots, N_T$:
**Step 2**: Observe uncertainty realizations $\widehat{W}_{t,m}$ from the historical data and obtain the decisions $x_{t,m}$ and the post-decision state $S_{t,m}^x$ by solving (26) and (27) based on the current slopes.
**For** $i = 1, \dots, N_I$:
**Step 3**: Calculate the observation of the marginal contribution of each $\widehat{SOC}_{i,t,m}$ to the $C_t^{obj}(S_{t,m}, x_{t,m})$ by (28).
**Step 4**: Obtain the changes of all post-decision states and then calculate the observation of the marginal contribution of each $\widehat{SOC}_{i,t,m}$ to the $S_{t,m}^x$ by (29).
**End** for ($i$);
**End** for ($t$);
/* **Backward Pass** */
**For** $t' = N_T, \dots, 1$:
**Step 5**: Update the slopes by **(30)**.
**Step 6**: Retain the convexity of PLFs by the concave adaptive value estimation algorithm.
**End** for ($t'$);
**End** for ($m$);
**Step 7**: Output all the resulted PLFs.

To update the sloped of PLFs, a sample observation of the marginal value $C_{i,t,m}^{\delta}(S_{t,m})$ and $R_{i,t,m}^{\delta}(S_{t,m}^x)$ is calculated as in (28) and (29) by imposing small perturbations to $\widehat{SOC}_{i,t,m}$.

$$C_{i,t,m}^{\delta}\left(S_{t,m}\right) = \partial C_t^{obj}(S_{t,m}, x_{t,m}) / \partial \widehat{SOC}_{i,t,m} \tag{28}$$

$$R_{i,t,m}^{\delta}\left(S_{t,m}^x\right) = \partial S_{t,m}^x / \partial \widehat{SOC}_{i,t,m} \tag{29}$$

Then, the slopes are updated by combining the current approximation and sample slopes via the step size $\lambda$:

$$r_{i,t-1,s,m} = (1-\lambda) \cdot r_{i,t-1,s,m-1} + \lambda \cdot r_{i,t,s,m}^{\delta} \tag{30}$$

where

$$r_{i,t,s,m}^{\delta} = \begin{cases} C_{i,t,m}^{\delta}\left(S_{t,m}\right) + R_{i,t,m}^{\delta}\left(S_{t,m}\right) \cdot r_{i,t+1,s,m}^{\delta}, & t \ne N_T \\ C_{i,t,m}^{\delta}\left(S_{t,m}\right), & t = N_T \end{cases}$$

Further, the concave adaptive value estimation algorithm [10] is adopted to retain the convexity of PLFs as below:

$$r_{i,t-1,s,m} = \begin{cases} \max\left\{r_{i,t-1,s,m-1}, (1-\lambda) \cdot r_{i,t-1,s,m-1} + \lambda \cdot r_{i,t,s,m}^{\delta}\right\}, s > s^* \\ (1-\lambda) \cdot r_{i,t-1,s,m-1} + \lambda \cdot r_{i,t,s,m}^{\delta}, s = s^* \\ \min\left\{r_{i,t-1,s,m-1}, (1-\lambda) \cdot r_{i,t-1,s,m-1} + \lambda \cdot r_{i,t,s,m}^{\delta}\right\}, s < s^* \end{cases} \tag{31}$$

where $s^*$ is the segment derived in $m$th iteration.

### D. Overall Application Process

In general, the proposed dimension-reduced ADP method consists of two parts: offline training and real-time operation. In the offline stage, the slopes of the PLF segments are trained using historical data until convergence. The resulting value function approximation, which embeds empirical information from historical operating conditions, is then used in real-time operation to support near-optimal decisions. Specifically, at each period $t$, the system operator first observes the current exogenous information and then determines the real-time control action by solving (26)-(27) with the pre-trained value function. Since (26)-(27) is a linear program, it can be solved efficiently using off-the-shelf commercial solvers.

## IV. Case Studies

This section presents three case studies to validate the effectiveness and scalability of the proposed method. All simulations are calculated using GUROBI 12.0.1 on a computer with a 2.30 GHz CPU and 16 GB RAM.

### A. Deterministic Case

The proposed method is first tested on the 33-bus system under a deterministic scenario. The real-time operation horizon is 24 h with a 15 min resolution. The system includes two 2 MW wind turbines, four 0.5 MW gas turbines, and four nodes each equipped with 60 air-conditioning loads, resulting in 240 units in total. The electricity price profiles are shown in Fig. 1. The parameter profiles are shown in Fig. 2, including one deterministic forecast (curves with markers) and multiple uncertainty scenarios (curves without markers). The remaining system parameters are available in [14].

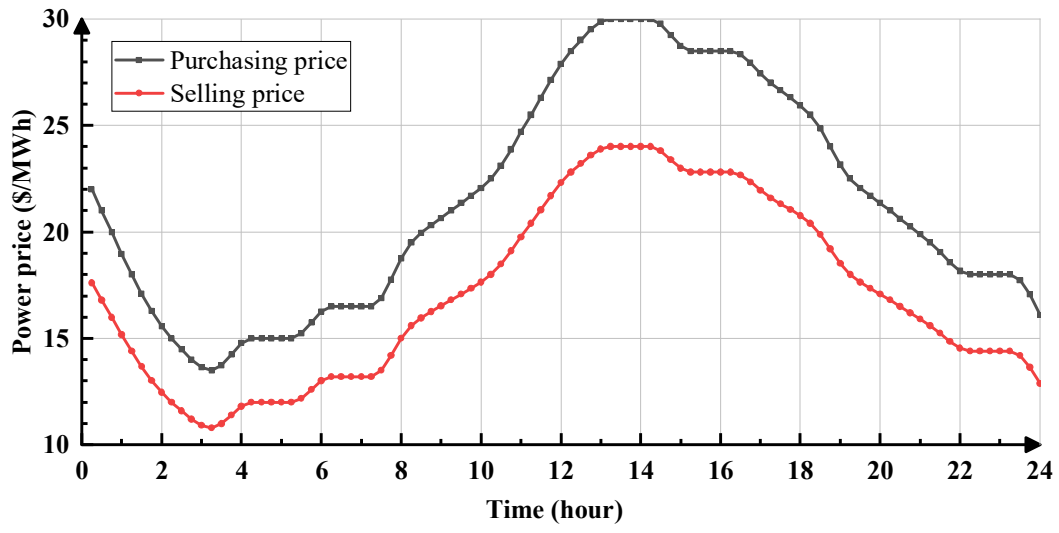

Fig. 1. Price profiles.

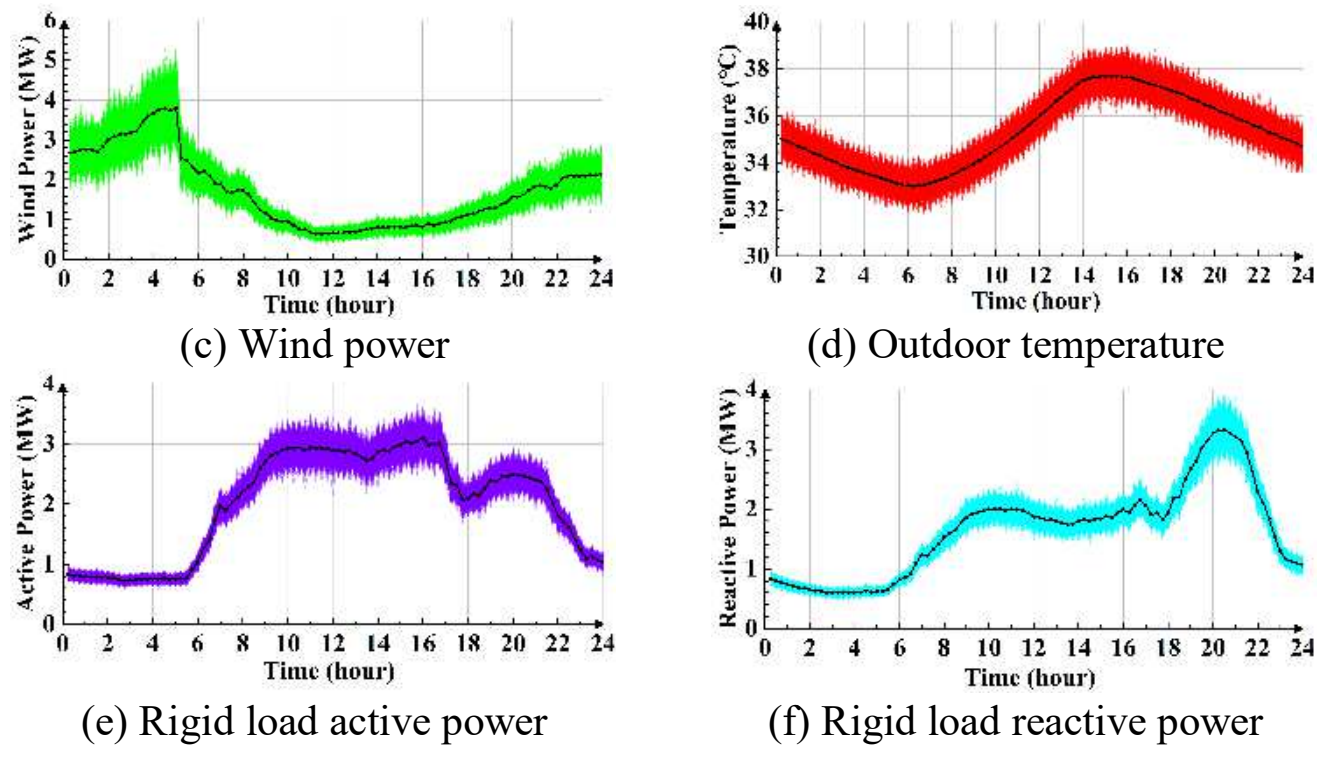

Fig. 2. Uncertain parameter curves.

Under these settings, the following methods are compared.

**Case 1**: **Oracle.** This method assumes perfect prediction and provides hindsight-based optimal results. Since it violates time causality and is infeasible in practice, it is used only as a lower bound for evaluation. The *solution gap* is defined as the absolute relative deviation of the total operating cost from that of the Oracle case.

**Case 2**: **Proposed dimension-reduced ADP.**

**Case 3**: **Myopic.** The real-time decision at period $t$ is obtained by solving a single-period optimization problem using only the information available at period $t$.

**Case 4**: **MPC.** The real-time decision at period $t$ is obtained by solving an $H$-period optimization problem over periods $t$ to $t+H$. Since future decisions in MPC depend heavily on forecast accuracy, $H$ is set to 8.

The results are presented in Table I and Figs. 3-4.

As shown in Table I, the myopic and MPC methods produce only locally optimal decisions, leading to solution gaps of 10.77% and 7.52%, respectively. By contrast, the proposed method achieves a near-optimal solution over the entire operation horizon, reducing the gap to 1.50%.

Fig. 3 shows that the SOC trajectory under the proposed method closely follows that of the Oracle method, indicating that the trained value function effectively captures the impact of current decisions on future operating costs. In contrast, the myopic and MPC methods fail to utilize the flexibility of air-conditioning loads efficiently. Since they rely only on current information or short-term forecasts, they cannot proactively adjust SOC to prepare for wind power accommodation during 4:00–5:00, which leads to wind curtailment.

As illustrated in Fig. 4, under Case 2, air-conditioning loads store cooling energy during 3:00–5:00 to absorb excess wind power. Meanwhile, the loads at buses 13 and 15 release cooling energy during 0:00–3:00 to increase upward regulation capability by leveraging building thermal inertia and preparing for the upcoming wind power accommodation. As a result, no wind curtailment occurs in Case 2. A similar advantage can be observed during the high-price purchasing period around 13:00–14:00. The myopic and MPC methods release cooling energy too early, forcing the microgrid to purchase more electricity from the external grid when prices are high. By contrast, under Case 2, air-conditioning loads release cooling energy during 11:45–19:45 to reduce demand during high-price periods. Although the purchasing price exceeds the response cost after 8:30, the loads do not respond immediately because the proposed method anticipates the even higher price around 13:00 and strategically preserves flexibility for that period. A similar pattern is also observed around 18:00, when the air-conditioning loads at buses 26 and 28 first store cooling energy and then release it to provide greater downward regulation capability.

Overall, these results show that the proposed method can effectively coordinate air-conditioning loads and achieve near-optimal economic operation.

TABLE I. PERFORMANCE IN DETERMINISTIC CASE

| Comparison terms | Oracle | Proposed ADP | MPC | myopic |
|---|---|---|---|---|
| Operation cost ($) | 700.5 | 711.0 | 753.1 | 775.9 |
| Solution gap (%) | 0 | 1.50 | 7.52 | 10.77 |
| Wind Curtailment (kWh) | 0 | 0 | 83.9 | 108.2 |

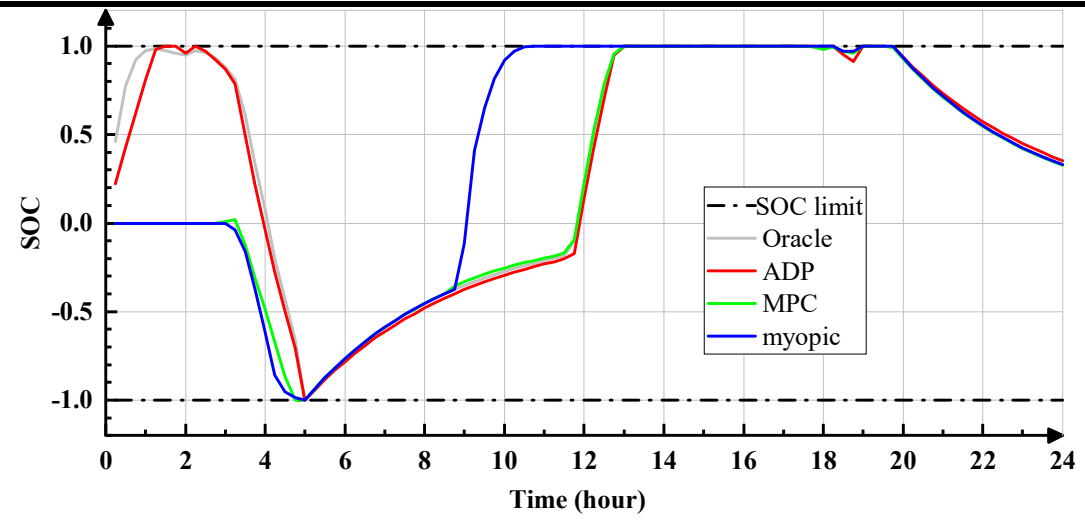

Fig. 3. Mean SOC profile at bus 13 under different methods.

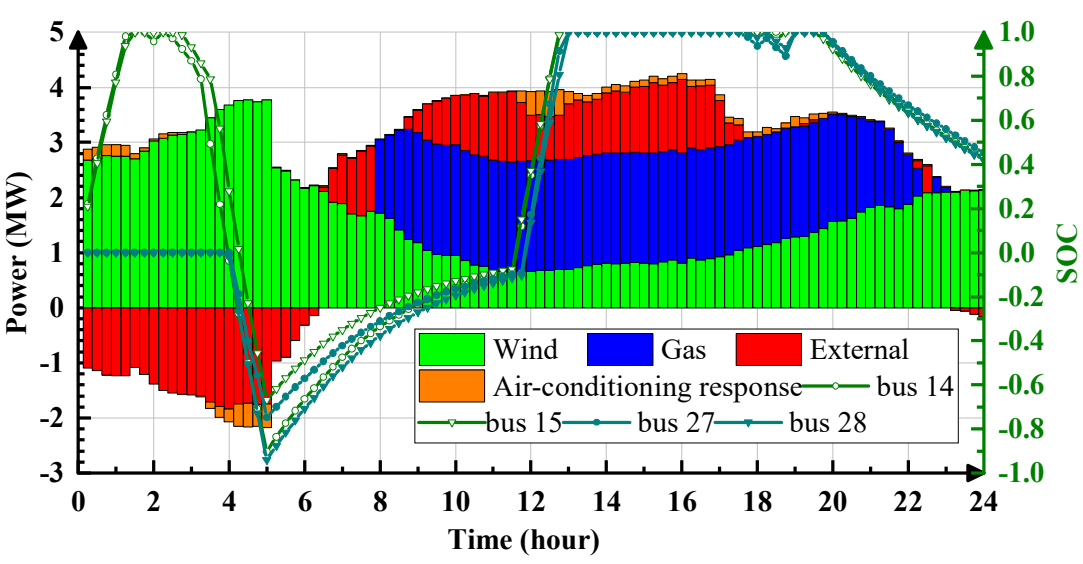

Fig. 4. Generation schedule in Case 2.

## *B. Stochastic Case*

Next, stochastic simulations are conducted using 2200 scenarios sampled from the historical data shown in Fig. 2. Among them, 2000 scenarios are used for offline value function training and the remaining 200 scenarios are used for real-time testing. The results are shown in Fig. 5.

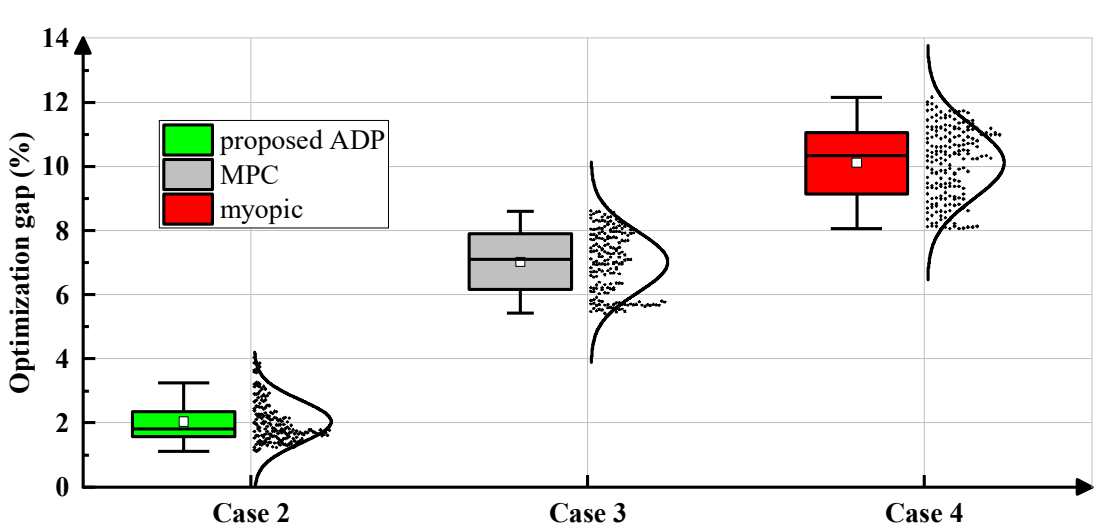

Fig. 5. Real-time testing results in the stochastic case.

The proposed ADP method achieves an average solution gap of about 2.10% over the 200 testing scenarios, significantly outperforming the other methods. In addition, its solution gap distribution is more concentrated, indicating stronger performance under multiple uncertainties. Since the value function is trained from historical data and captures the distribution of prediction errors, the proposed ADP method determines the real-time control action based only on the currently observed information. In this sense, it is similar to the myopic method, but achieves a much smaller solution gap. These results further verify that the proposed method can effectively support economic real-time microgrid operation under multiple uncertainties.

### C. Scalability Test

The scalability and computational efficiency of the proposed method are further evaluated on the modified IEEE 123-bus system. The system includes ten nodes each equipped with 50 air-conditioning loads, resulting in 500 units in total. The parameter profiles are similar to those in Fig. 2 and are scaled to match the larger system. The numbers of training and testing scenarios are set to 2000 and 200, respectively. For comparison, an additional case based on the traditional ADP method without dimension reduction is also included. The results are shown in Fig. 6 and Table II.

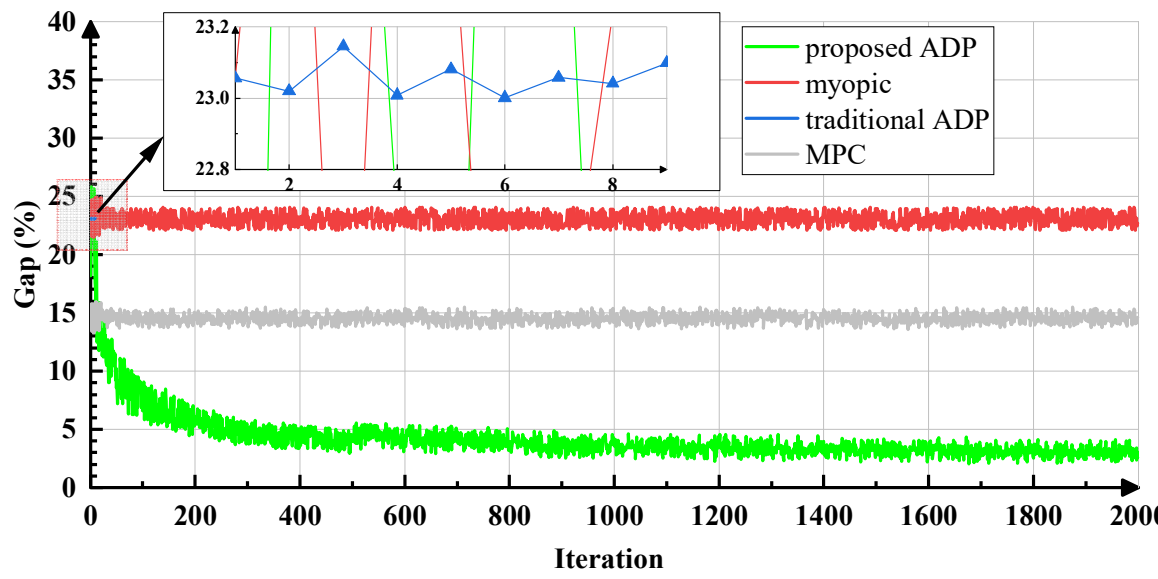

Fig. 6. Offline training results in scalability test.

TABLE II. REAL-TIME TESTING RESULTS IN SCALABILITY TEST

| Comparison terms | Oracle | Proposed ADP | MPC | myopic |
|---|---|---|---|---|
| Average gap (%) | 0 | 2.87 | 14.63 | 23.17 |
| Average CPU time (s) | 10.53 | 2.27 | 21.09 | 7.94 |

As shown in Fig. 6, the proposed ADP method converges rapidly and reaches the smallest training gap, about 3%, after approximately 1000 iterations. By contrast, the traditional ADP method without value function projection requires about 9400 s per iteration because of the extremely large state space, allowing only nine updates within one day. This is insufficient for effective value function training. The proposed method, by comparison, requires only about 16 s per iteration. Moreover, the traditional ADP method converges more slowly because a large number of value function slopes must be updated, making it impractical for real applications.

Table II further shows that the proposed ADP method achieves the smallest real-time testing gap among all practical methods because it effectively exploits the empirical knowledge learned from historical data. It also requires the shortest computation time, benefiting from the proposed dimension reduction method. These results demonstrate the computational tractability, solution quality, and scalability of the proposed method in larger-scale systems.

## V. Conclusion

This paper proposes a dimension-reduced ADP method for real-time microgrid operation with massive air-conditioning loads under multiple uncertainties. By combining post-decision value functions with a consistency-based state projection, the proposed method makes sequential decision-making with large numbers of flexible loads computationally tractable. The case studies show that the method can achieve a good balance among solution quality, computational efficiency, and scalability under stochastic conditions.

From a practical perspective, the proposed method can provide microgrid operators with an efficient and scalable energy management framework for utilizing the regulation potential of massive air-conditioning loads in real time. One limitation of this work is that the set of air-conditioning loads participating in demand response is assumed to be determined day-ahead. How to handle time-varying participation of air-conditioning loads in real-time operation is left for future work.